\documentclass[aps,
               jap,
               reprint,
               amsfonts,
               unsortedaddress,
               superscriptaddress,
               amssymb,
               showkeys]{revtex4-1}

\usepackage{siunitx}

\usepackage{amsmath}      
\usepackage[dvips]{graphicx}     
\usepackage{color}        
\usepackage{hyperref}     

\newcommand{\CAU}{Christian-Albrechts-Universit\"at zu Kiel}

\newcommand{\ITAP}{Institut f\"ur Theoretische Physik und Astrophysik, \CAU, Leibnizstra{\ss}e 15, D-24098 Kiel, Germany}
\newcommand{\MAWI}{Institut f\"ur Materialwissenschaft, Lehrstuhl f\"ur Materialverbunde, \CAU, Kaiserstra{\ss}e 2, D-24143 Kiel, Germany}

\newcommand{\MAILBONITZ}{bonitz@theo-physik.uni-kiel.de}
\newcommand{\INST}[1]{\affiliation{#1}}

\usepackage{xcolor}

\begin{document}

\preprint{AIP/123-QED}

\title[MD simulations of Au cluster growth]
{Molecular dynamics simulation of gold cluster growth during sputter deposition}

\author{J. W. Abraham}
\INST{\ITAP}


\author{T. Strunskus}
\INST{\MAWI}

\author{F. Faupel}
\INST{\MAWI}

\author{M. Bonitz}
\email[mail to: ]{\MAILBONITZ}
\INST{\ITAP}

\date{\today}

\begin{abstract}
We present a molecular dynamics simulation scheme
that we apply to
study the time evolution of the
self-organized growth process
of metal cluster assemblies formed by sputter-deposited
gold atoms on a planar surface.
The simulation model incorporates
the characteristics of the plasma-assisted deposition
process and allows for an investigation
over a wide range of deposition parameters.
It is used to obtain
data for the cluster properties
which can directly be compared to
recently published experimental data for gold
on polystyrene
(M. Schwartzkopf \textit{et al}., ACS
Appl. Mater. Interfaces \textbf{7}, 13547 (2015)).
While good agreement is found between the two,
the simulations additionally provide valuable
time-dependent real-space data
of the surface morphology some of whose
details are hidden in the reciprocal-space 
scattering images that were used
for the experimental analysis.
\end{abstract}

\maketitle

\section{Introduction}
Nanocomposites
have been subject to extensive experimental and theoretical
research in the last decades because the assembly
of materials with contrasting properties may result
in remarkable characteristics of the composite.
The variety of technological applications based on nanocomposites
can be found in, e.\,g., electronics \cite{Fulton87, Graf02, Takele08, Karttunen08}, 
plasmonics \cite{Biswas04, BiswasAppl04}, food
packaging \cite{Emamifar11} and medicine \cite{Fan02, Faupel2007}. While various types 
of materials are present in synthetic and natural nanocomposites, 
metal-polymer nanocomposites are of specific interest
as they are promising candidates for the
fabrication of materials with tailored magnetic, electronic and optical properties 
\cite{Pomogailo2006, Ramesh2009, n9_wang, springerbook14, Schwartzkopf2013, Greve2006, rosenthal13}
at low cost. 

Typical physical vapor deposition processes of metal-polymer nanocomposites 
comprise simultaneous deposition of metal and polymer,
leading to nanoparticles embedded in a polymer
host matrix, or exclusive deposition of metal atoms or clusters onto a prepared polymeric
thin film, restricting the nanoparticle formation to the surface and near-surface
area of the polymer bulk \cite{rosenthal11, Greve2006, zaporojtchenko2000formation, Bechtolsheim1999, Schwartzkopf2015}.
Depending on the desired properties of the composite, different
deposition methods of
both polymer and metal may be advantageous \cite{Faupel2010, zaporojtchenko2000formation, schuermann2005, Depla2008, smith1995thin}.
For example, sputter deposition of metal atoms allows for high deposition rates,
but the employed plasma also affects the surface structure (see below).
While many of the experimental approaches have the common goal
of getting independent control over 
sizes, shapes and spatial distributions of
the nanoparticles, they also share the difficulty
that the self-organized nanoparticle formation process can only
be influenced indirectly, e.\,g., by changing deposition rates
or substituting materials.
Therefore, in situ observations
of the film characteristics during the deposition process
are useful to deepen the understanding of the relevant growth kinetics.

Recent progress in experimental monitoring of film growth
was made by applying
grazing incidence small-angle
X-ray scattering (GISAXS)
to trace the time-resolved morphology of sputter-deposited gold on polymer surfaces.
\cite{Metwalli2008, Schwartzkopf2013, Schwartzkopf2015}.
This experimental method relies on extracting all structural
information from the features in the scattering patterns.
GISAXS experiments are well suited to measure the evolution
of mean cluster properties such as radii, heights and distances
between clusters, but they lack the possibility to observe the atomic structure of
individual clusters in real space.
Furthermore, it is impossible to capture
the microscopic physical processes that drive the cluster
formation.
These processes comprise
thermally activated diffusion of metal atoms on the surface,
desorption of single atoms with high kinetic energy,
coalescence of metal clusters, and direct attachment
of deposited atoms to existing clusters.
At that, the interaction between metal atoms is typically
much stronger than
that between metal atoms and the polymer.
Furthermore, the plasma which is
applied to sputter atoms from the metallic
target also influences the cluster growth.
Not only does it transfer
charges to the nanoparticles, by ejecting highly
energetic ions in the direction of the substrate it also
causes defects at the surface, where the trapping probability
of metal atoms is strongly increased \cite{smirnov2009cluster, mattox1989}.

One possibility to improve the understanding of the complex
interplay of the mentioned processes is offered by computer simulations.
So far, however, we are not aware of any computational studies of
thin-film growth by sputter deposition which resolve not only the atomic structure of the clusters,
but also take into account the influence of the polymer and the plasma background.
Although kinetic Monte Carlo simulations have proved to be
appropriate for the description
of similar systems,
they are usually restricted to cases with simple cluster geometries
that allow one to neglect the motion of individual atoms
such that clusters can be approximated by simple geometrical objects, e.\,g., spheres or columnar
structures \cite{rosenthal13, springerbookchapter, Abraham15}.
In this work, we go beyond these
limitations by introducing an atomic scale molecular dynamics (MD) simulation scheme
for the motion of sputter-deposited gold particles onto a polymeric surface.
Although the incorporation of individual atoms is on the cost of accessible
system sizes,
the method allows us to perform computations with surface sizes of at least $\SI{45}{nm} \times \SI{45}{nm} $.
As we will show in a parameter study for film thicknesses up to \SI{3}{nm}, the simulated
systems are big enough to directly compare and find agreement with
recently published experimental data for sputtered
gold (Au) on a polystyrene (PS) film~\cite{Schwartzkopf2015}. Furthermore, the employed parameters
and potentials can easily be adjusted in order to represent
different plasma conditions or materials.

Although
molecular dynamics has been a successful and established method
to simulate the deposition of thin films 
for several decades, e.\,g., \cite{Mueller1987, Gilmore1991, Haberland1995, Dong1996, Rozas07, Xie2014},
our work attains distinction through
the incorporation of the characteristics of sputter-deposited thin metal films
as well as the possibility to perform close comparisons with experimental data.
In the following, we briefly 
explain in what aspects our simulation method differs from other common approaches.

First, the treatment of the polymer surface is not particle-based.
Instead the diffusive motion of the metal atoms on the surface
is driven by fluctuating and dissipative forces.
This is much simpler than
atomistic or coarse-grained surface models,
but it has been shown in previous kinetic Monte Carlo (KMC) studies \citep{Thran1997,rosenthal11,bonitz12,rosenthal13,Abraham15}
that the diffusive motion of the metal atoms can be appropriately described
by employing a continuum model for the polymer.
This treatment is physically motivated by the facts
that the assembly of polymer chains is very disordered and
the interaction between metal and polymer is typically very weak \cite{Faupel1998}.
The stochastic description of the atomic motion allows us to
set diffusion coefficients and spend the saved
computer resources on the simulation of more metal atoms.
For example, the number of particles in our simulations
is about two orders of magnitude larger than typical numbers
in simulations with particle-based substrates \cite{Rozas07, Xie2014}.
Another advantage of a tunable
diffusion coefficient
is that it is straightforward to obtain realistic values of the
distance that a diffusing atom travels during the average
time between the deposition of two particles. In a particle-based model of the polymer surface,
this would require a more complicated modification
of the heat bath or the potentials of the cross interaction
between metal and polymer particles.

As a second difference, we implemented
some 
idealized model processes in order to
mimic the re-evaporation of atoms
from the surface and the creation of surface defects.
Although these processes do not represent
the exact atomistic behavior of the system,
they enable us to make statements about the qualitative
influence of the corresponding real physical processes,
and they offer the advantage of being easily adjustable
to specific conditions.
Hence, we combine the advantages of a microscopic treatment, which is applied in
standard molecular dynamics, and the introduction of simplified model processes,
which is typical for kinetic Monte Carlo simulations \cite{Jansen2012}.

Further details
of the simulation scheme and its relation to realistic experimental
scenarios are explained in Sec.~\ref{sec:simulationscheme}.
Following this, in Sec.~\ref{sec:results}, a comparison with experimental data 
and a parameter study of cluster properties
for various trapping parameters, re-evaporation rates and deposition rates are presented.

\begin{figure}
\begin{center}
\includegraphics[scale=0.93]{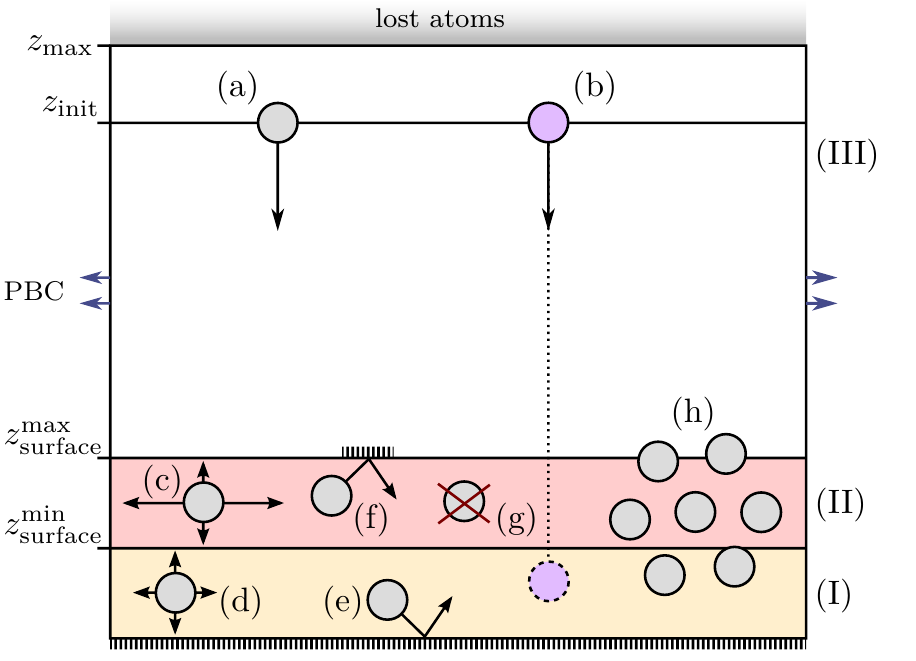}
\end{center}
\vspace{-0.2cm}
\caption{Illustration of atomic processes in different parts
of the simulation box (side view):
deposition of atoms (a), deposition and implantation of ions (b), diffusion in the surface layer (c)
and in the bulk layer (d), reflection of atoms at the bottom wall of the simulation box (e),
reflection of neighborless atoms at the top of the surface (f),
removal of atoms from the surface due to re-evaporation (g),
cluster formation and crossing of the border at $z_\mathrm{surface}^\mathrm{max}$ (h).
The layers correspond
to the polymer bulk (I), the surface (II) and the plasma
environment above the surface (III).
While the simulation box has periodic boundary conditions (PBC) in horizontal directions,
atoms that dissociate from clusters 
and small clusters that very rarely dissociate from the surface are removed from
the simulation after crossing the top border of the simulation box.
}
\label{fig:simulationbox}
\end{figure}

\section{Simulation scheme}
\label{sec:simulationscheme}
All simulations were carried out using the \textit{LAMMPS} software package \cite{lammps}.
The code performs standard molecular dynamics simulations with specific boundary
conditions and additional processes for the addition and removal of atoms.
For all particles, the interatomic interaction potential is calculated with the embedded-atom method \cite{daw84},
using tabulated data for gold provided by the work of Foiles \textit{at al.} \cite{foiles86}.
In the following, we describe the partitioning of the simulation box and the 
specific treatment of the particles.

\subsection{Details of the model}
As illustrated in Fig.~\ref{fig:simulationbox}, the simulation box with the dimensions
$ [0, x_\mathrm{max}]$, $ [0, y_\mathrm{max}]$, $ [0, z_\mathrm{max}]$
and periodic boundary conditions in $x$-$y$-directions is divided into different layers
that are intended to represent the behavior of the particles above the
surface ($z > z_\mathrm{surface}^\mathrm{max}$), on the surface ($z_\mathrm{surface}^\mathrm{min} < z \leq z_\mathrm{surface}^\mathrm{max}$) and in the uppermost part of the polymer bulk
($0 < z \leq z_\mathrm{surface}^\mathrm{min}$).
We stress that the surface layer has a finite thickness in order
to take into account the roughness of the polymer surface.

For all simulation results presented in this work, we used the values
$x_\mathrm{max}=y_\mathrm{max}= \SI{45}{nm}$,
$z_\mathrm{max}= \SI{8}{nm}$, $z_\mathrm{surface}^\mathrm{min} = \SI{0.2}{nm}$
and $z_\mathrm{surface}^\mathrm{max} = \SI{0.6}{nm}$.
As mentioned before,
the dynamics of individual building blocks of the polymer chains are ignored in this model.
Instead, the polymer is treated as a continuous fluid-like background to the gold
particles, enabling us to perform Langevin dynamics simulations.
In the subsequent paragraphs, we address further details of the simulation process.
\paragraph{Iteration cycle}
All quantities of interest are monitored after each iteration period consisting of
$N_\mathrm{iter}=300$
steps.
This involves updates of the associated layers and calculations of the coordination number
to identify atoms on the surface without neighbors.
\paragraph{Creation of particles}
It is not necessary to simulate the processes in the Au cathode
that lead to the ejection of atoms. Instead, we assume that there is a constant
flux of sputtered atoms and ions that travel towards the surface. We model this by continually creating
atoms and ions at random positions in the plane at $z_\mathrm{init}=\SI{7}{nm}$
(cf.~Fig.~\ref{fig:simulationbox}) with the initial
velocity $\mathbf{v}_\mathrm{init} = \left(0, 0, -|v_\mathrm{init}| = \SI{-0.1}{nm/ps} \right)$
\footnote{
Depending on the plasma parameters and thermal broadening, the deposited particles may have different
kinetic energies. This only has a weak influence as long as no resputtering occurs
or defects in the polymer are created. The latter effect is taken into account by a separate process in the model.}.
The time between
particle creation events can be calculated from 
the deposition rate $J$.

\paragraph{Motion of the particles}
With the exception of the space between $z_\mathrm{surface}^\mathrm{max}$ and
$z_\mathrm{max}$, where the motion of the metal particles is treated microcanonically,
the diffusive motion on the surface is modeled by friction and fluctuating forces using
Langevin dynamics \cite{Schneider1978}. 
Thus we assume that the polymer acts like a continuous heat bath in thermal
equilibrium.
The corresponding equation of motion for all particles with the coordinates
$\mathbf{r} = \left( \mathbf{r}_1, \mathbf{r}_2, \dots \right) $ reads
\begin{equation}
m \ddot{\mathbf{r}} = - \nabla U(\mathbf{r})  -\frac{m}{t_\mathrm{damp}} \dot{\mathbf{r}}
+ \mathbf{F}_\mathrm{fluct} \;,
\end{equation}
where $m$ is the mass of a gold atom, $t_\mathrm{damp}$
controls the strength of the friction, and $U$ is the interatomic potential.
The fluctuating force obeys the proportionality
\begin{equation}
\mathbf{F}_\mathrm{fluct} \propto  \sqrt{ \frac{k_\mathrm{B}Tm }{ t_\mathrm{damp} \Delta t } } \;,
\end{equation}
where $T$ is the substrate temperature, $k_\mathrm{B}$ is the Boltzmann constant,
and $ \Delta t$ is the timestep.
The direction and the magnitude of $\mathbf{F}_\mathrm{fluct}$
are randomized according to Ref.~\cite{Dunweg91}.
\paragraph{Diffusion}
Metal atoms on polymer substrates are known to not only diffuse on the surface,
but also to penetrate slowly into the bulk. 
The application of Langevin dynamics allows us to simulate such an anisotropic
diffusive motion by adjusting the damping parameters $t_\mathrm{damp}$
for specific directions and areas of the simulation box.
The corresponding diffusion coefficients can be calculated via
\begin{equation}
\label{eq:diffusioncoeff}
D=  \frac{1}{m} {k_\mathrm{B}T t_\mathrm{damp}} \;.
\end{equation}
In the surface layer,
the diffusion coefficient 
for the horizontal motion, $D_\mathrm{surface}^\parallel$,
is assigned a value that is bigger than the
one for the vertical motion, 
$D_\mathrm{surface}^\perp$.
For the bulk layer, we set $D^{x/y/z}_\mathrm{bulk}=D_\mathrm{surface}^\perp$,
i.\,e., we keep the diffusion coefficient
of the vertical motion in the surface layer and assign it to all directions.
As it is stated in Ref.~\cite{Thran1997}
that
the ratio
\begin{equation}
\alpha = \frac{D_\mathrm{surface}^\parallel}{D^{x/y/z}_\mathrm{bulk}} \;.
\end{equation}
is presumably larger than 60, we performed all
simulations with $\alpha=80$.
This choice is rather arbitrary, 
but we carefully verified that the results do not show
significant differences for ratios as low as $\alpha=30$.
Furthermore, the value can easily be adjusted once it has been determined for specific
materials in external studies.
In addition to the reduction of the bulk diffusion by decreasing the corresponding
diffusion coefficients $D^{x/y/z}_\mathrm{bulk}$ and $D_\mathrm{surface}^\perp$,
all atoms are reflected
at the bottom of the simulation box.
This is motivated by the fact that
deep penetration into the bulk is not
observed in the experiments from Ref.~\cite{Schwartzkopf2015}.
However, we remark that there exist other experimental
situations which require an accurate description
of clusters diffusing in the polymer bulk \cite{Bechtolsheim1999}.

\paragraph{Re-evaporation} 
The condensation coefficient $C$ of metal on metal
is close to one due to the strong metallic bonds.
Hence, re-evaporation of atoms from clusters is rarely
observed in experiments. In the simulations, the
employed potential ensures that this behavior is reproduced.
In contrast, the condensation coefficient of metals on polymers at room temperature
strongly depends on the involved materials.
For example, the values $C=0.006$, for Au on Teflon AF, and  $C=0.955$, for
Au on PMDA-ODA polyimide, were reported
\cite{Zaporojtchenko2000, Zaporojtchenko1999}.
Consequently, re-evaporation of metal atoms from the surface has to be considered
in the simulations.
In fact, if we performed pure 
Langevin dynamics in the surface layer,
some free atoms would occasionally cross the border at $z_\mathrm{surface}^\mathrm{max}$
and then be likely to reach the top of the simulation box and get lost.
However, this behavior does not correspond to the actual physical
process and it cannot be controlled without changing other
important model parameters.
We therefore implement another simple model process that allows
us to take into account the desorption of single metal atoms.
For that purpose,
we force all free metal atoms to remain in the two bottom layers by
defining another reflective boundary
at $z_\mathrm{surface}^\mathrm{max}$.
This boundary only acts on atoms that
have no neighbors within the cutoff distance of the interaction potential,
i.\,e., atoms belonging to a cluster are 
not affected by this modification.
In a next step, we define a process that removes
neighborless atoms from the surface layer with a probability
$p_\mathrm{re}$ within each iteration cycle.
This probability 
can be understood as a parameter that mimics the characteristic sticking behavior
of a specific material.

\paragraph{Lost atoms}
Atoms that belong to clusters ($N_\mathrm{atoms}>1$) are
allowed to cross the boundary at $z_\mathrm{surface}^\mathrm{max}$.
Due to the temperature-induced fluctuating forces, there is a very small, but finite
probability that atoms escape from a cluster. Most of the time,
these atoms travel in a straight line
towards the top of the simulation box and are removed once they reach $z_\mathrm{max}$.

\paragraph{Trapping of atoms}
A certain fraction of the
gold atoms adheres to the surface
immediately or shortly after they reach the
surface in an actual experimental set-up.
There are several reasons for this behavior.
On the one hand, the untreated polymer surface
exhibits some sites with increased trapping
probabilities due to its unevenness and imperfections.
Furthermore, during the deposition process,
additional surface defects
are created by the impingement of highly energetic
ions from the plasma, e.\,g., Ar$^+$ if an Argon plasma is used.
On the other hand, knowing that 
typical energy distributions of sputtered
metal particles have a long tail that exceeds
several tens of eV \cite{Stuart1969},
one can expect
that some of the sputtered Au atoms
are sufficiently fast for being implanted
into the polymer bulk.
Since there do not exist detailed microscopical
studies of these effects, neither the sizes
nor the number of surface defects are known.
For that reason, we restrict ourselves to
simulating the qualitative behavior 
of the surface defects
with
a specific treatment for a fraction $\gamma$
of all deposited particles. 
If one of these particles reaches a
point below $z_\mathrm{surface}^\mathrm{min}$,
its position is kept fixed, but 
the interaction with the other atoms is maintained.
If instead the particle is deposited onto
other metal atoms and thus does not reach the surface layer,
it is treated just like the other atoms.

\begin{figure}
\begin{center}
\includegraphics[scale=1.00]{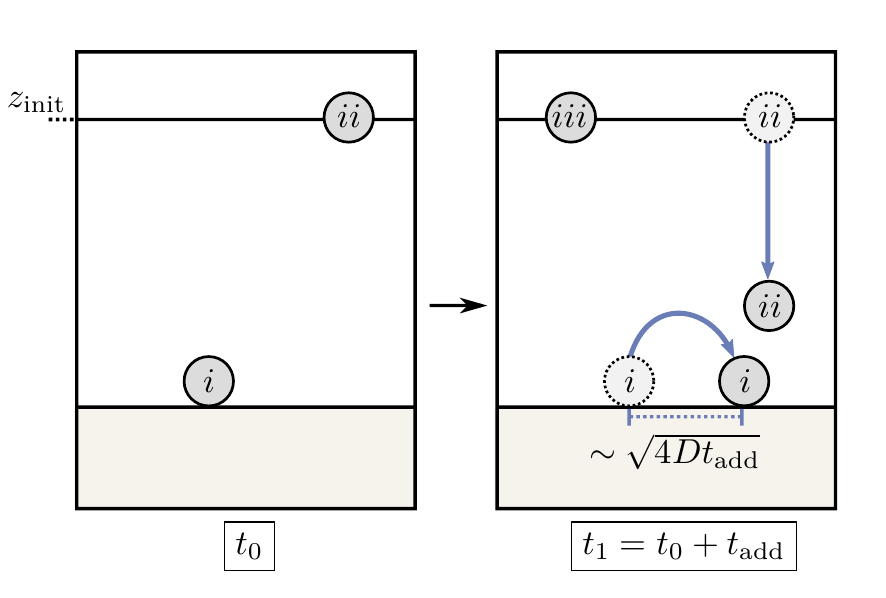}
\end{center}
\vspace{-0.5cm}
\caption{Principle of the proportional rescaling 
of the deposition rate $J$ and the diffusion coefficient $D$.
For all combinations of $J$ and $D$ with $J/D=\mathrm{const.}$,
the mean squared displacement of particle $i$ is preserved
during the time $t_\mathrm{add}$
that it takes to create particles $ii$ and $iii$
at random positions in the plane at $z_\mathrm{init}$.
}
\label{fig:meansquareddispl}
\end{figure}

\paragraph{Adjustment to experimental parameters}
In the experimental set-up presented in Ref.~\cite{Schwartzkopf2015},
the deposition rate $J_\mathrm{exp}=\SI{0.49}{nm/min}$
and the surface diffusion coefficient $D_\mathrm{exp}=\SI{7.33e-18}{m^2/s}$
were measured.
From the knowledge of $J_\mathrm{exp}$, one can calculate
the average time between the deposition of two gold
atoms per \SI{}{nm^2}, which is about \SI{2}{s}. 
At the same time,  the mean squared displacement of normal two-dimensional
diffusive motion can be used to estimate the distance that an atom travels
during the time $t$, according to
\begin{equation}
\label{eq:travelleddistance}
l := \sqrt{\left( r(t) - r(0)  \right)^2 } = \sqrt{4 D_\mathrm{exp} t } \;.
\end{equation}
For $l=\SI{1}{nm}$, we obtain $t=\SI{0.03}{s}$.
These values indicate that the deposition process and the diffusive
motion of the particles are many orders of magnitude slower
than the  cluster formation processes, which happen on time scales
of pico- and nanoseconds. In order to capture
these fast processes, we set the timestep to
$\Delta t = \SI{0.001}{ps}$.
As the number of timesteps to obtain a sufficiently large system
must not exceed a practical limit,
this choice requires
that the simulations are performed with 
very high deposition rates $J$ and diffusion coefficients $D_\mathrm{surface}^\parallel$,
which considerably outreach typical experimental values.
This is a common problem in MD simulations of deposition processes.
For example, in Ref.~\cite{Gilmore1991} thermal activated diffusion
during the time between arriving atoms is neglected.
Another approach is to switch between NVE and NVT ensembles
to make sure that the system relaxes after each deposition event
\cite{Georgieva2009}.
Since the present model treats diffusion via Langevin
dynamics, it is possible to preserve
a realistic value for the mean squared displacement
of neighborless particles during the time between two deposition events.
This can be done
by performing
the simulations with proportionally increased experimental values 
of the deposition rate and the diffusion coefficient.
Hence, by maintaining the ratio
\begin{equation}
\label{eq:ratio_xi}
\xi := \frac{D^\parallel_\mathrm{surface}}{D_\mathrm{exp}} = \frac{J}{J_\mathrm{exp}}
\end{equation}
in the simulations,
the mean squared displacement of a diffusing particle, $l^2 \propto Dt$,
is preserved because both the diffusion coefficient and the time enter linearly.
An illustration of this principle can be found in Fig.~\ref{fig:meansquareddispl},
which displays an exemplary motion of a 
single particle on the surface during the time
$t_\mathrm{add} \propto J^{-1}$ between the deposition of two other particles.
In practice, we first define $D^\parallel_\mathrm{surface}$ according to Eq.~\eqref{eq:diffusioncoeff}
by setting the substrate temperature to the experimental value $T=\SI{296}{K}$
and choosing the value $t^\mathrm{surface, \parallel}_\mathrm{damp} = \SI{1}{ps}$
for the damping parameter. Afterwards, we take the values for
$D_\mathrm{exp}$ and $J_\mathrm{exp}$
from Ref.~\cite{Schwartzkopf2015} to calculate
the ratio $\xi$ and finally the deposition rate $J$.
For the mentioned parameters, we obtain the values
\SI{1.2}{ns} for the average time between the deposition
of two atoms per \SI{}{nm^2} and \SI{20}{ps}
for the time a diffusing atom to reach
a mean squared displacement of \SI{1}{nm^2}.

We remark that the employment of Langevin dynamics, which
treats all atoms on the surface and in the bulk
as if they are surrounded by a continuous fluid,
is only an approximation of the realistic diffusion process,
which is mainly caused by the interaction between the fluctuating
molecules of the polymer chain and the atoms at the outer cluster shell.
This is also the reason for the fact that one cannot make the simulations
more realistic just by decreasing the damping
parameter  $t^\mathrm{surface,\parallel}_\mathrm{damp}$
because that would lead to strongly damped 
motion of all cluster atoms that are in the bulk
or the surface layer. Although it is very difficult to
associate our choice of  $t^\mathrm{surface,\parallel}_\mathrm{damp}$
with specific materials, we made sure that 
the chosen parameter set enables fast formation, coalescence
and equilibration of stable clusters in the simulations.
Furthermore, we checked that despite the strongly increased
deposition rate, nucleation does not happen in the gas phase,
but only on the surface.

\section{Results}
\label{sec:results}
Before we present the simulation results in this section,
we mention some details about the evaluation of the cluster properties.
Furthermore, in order to show the limitations of the comparability between simulated
and experimental results,
we briefly explain how the experimental data
were obtained in Ref.~\cite{Schwartzkopf2015}.

\subsection{Evaluation of simulation results}
\label{sec:eval_sim_data}
In the following, we explain how we determine the cluster properties
from the atom positions obtained in the simulations. 

The first step of the evaluation is the identification of clusters.
For that purpose,
an atom is defined to belong to a cluster if its
distance to at least one of the cluster atoms
is below the threshold value \SI{0.32}{nm}.
The employed
algorithm was taken from the software \textit{OVITO} \cite{Stukowski2010Ovito}.
Furthermore, we remark that we define
that the minimum size of a cluster is 2 atoms, i.\,e.,
all neighborless atoms
on or above the surface are excluded from the evaluation
of the following quantities.

All presented cluster radii refer to the 
maximum of the horizontal distances of 
the cluster atoms and the center of mass of the cluster.
In order to get reasonable values in the case of clusters that extend over
one or more periodic images of the simulation box,
we use the generalized method for the calculation of the center of mass
in a system with periodic boundary conditions from Ref.~\cite{Bai2008}.
Of course, this definition of the radius is only a rough measure
for the cluster extension in the case of clusters with complex, non-spherical
shapes.

The cluster heights are characterized with the help of the
distribution function of vertical atom positions, $f(z)$. In the results provided below,
the heights are indicated by the values $\tilde{z}$ for which the cumulative distribution function,
\begin{equation}
\int_0^{\tilde{z}} f(z) \, \mathrm{d}z \;,
\end{equation}
reaches $0.99$, i.\,e., 99\,\% of all atoms have a vertical position below $\tilde{z}$.

For the mean cluster distance $D$, we use the estimator 
\begin{equation}
\label{eq:meanClusterDistance}
D = (A/N_\mathrm{cluster})^{1/2} \,
\end{equation}
where $A$ is the surface area of the
simulation  box and $N_\mathrm{cluster}$ is the total number of clusters
on the surface. Similar to the cluster radius, this quantity is
only meaningful as long as there is a large number of isolated homogeneously
distributed clusters on the surface.

In addition to these cluster properties,
we also provide results for the fraction of the surface that is covered
with metal. After binning the atom position on a two-dimensional
grid which is parallel to the surface,
the coverage is obtained by calculating the ratio
of the number of occupied bins and the total number of bins.
The side length of a bin is set to half the value of the lattice constant of Au.
While maintaining a high resolution,
this choice is big enough such that a
homogeneous gold layer corresponds to a surface coverage of one.

Finally, we remark that
we plot all quantities in dependence of the effective film thickness
$\delta$ which is the height of a homogeneous metal film that contains as many
atoms as can be found on the surface. 
We stress that this quantity is not
necessarily proportional to the time because due to desorption effects, the number
of atoms on the surface is usually less than the total number of deposited atoms.

\subsection{Experimental method}
\label{sec:exp_method}
Whereas the simulations provide atom positions in real space,
in Ref.~\cite{Schwartzkopf2015}, the surface morphology was
investigated in reciprocal space,
using the GISAXS method \cite{Levine1989,Levine1991}.
This method relies on the analysis of the angle-dependent intensity structure
of scattered X-rays. Most cluster properties were obtained by
projecting the scattered structures onto a geometrical model
of uniform hemispherical clusters distributed on a two-dimensional hexagonal lattice.
The cluster distance $D$ was calculated 
according to $D=2 \pi/q_0$, where  $q_0$
is the inverse length coordinate of the first lateral
maximum of the scattered intensity.
Similarly, the effective
thickness of the film
$\delta$
was determined from the vertical
structure of the scattered intensity.
Then, the cluster
radius was calculated such that the volume
of the hemispherical clusters is equal to
the effective volume of a homogeneous layer with the thickness
$\delta$. The cluster heights were independently
determined by fitting a hemispherical
cluster model to the measured GISAXS data
with the software \textit{IsGISAXS} \cite{Lazzari2002}.
Hence, when comparing the simulated data to the experimental data,
one has to keep in mind that some details of the local
film structure vanish by the averaging procedure
of generating a fit to simplified geometrical objects.

\begin{figure}
\begin{center}
\includegraphics[scale=1]{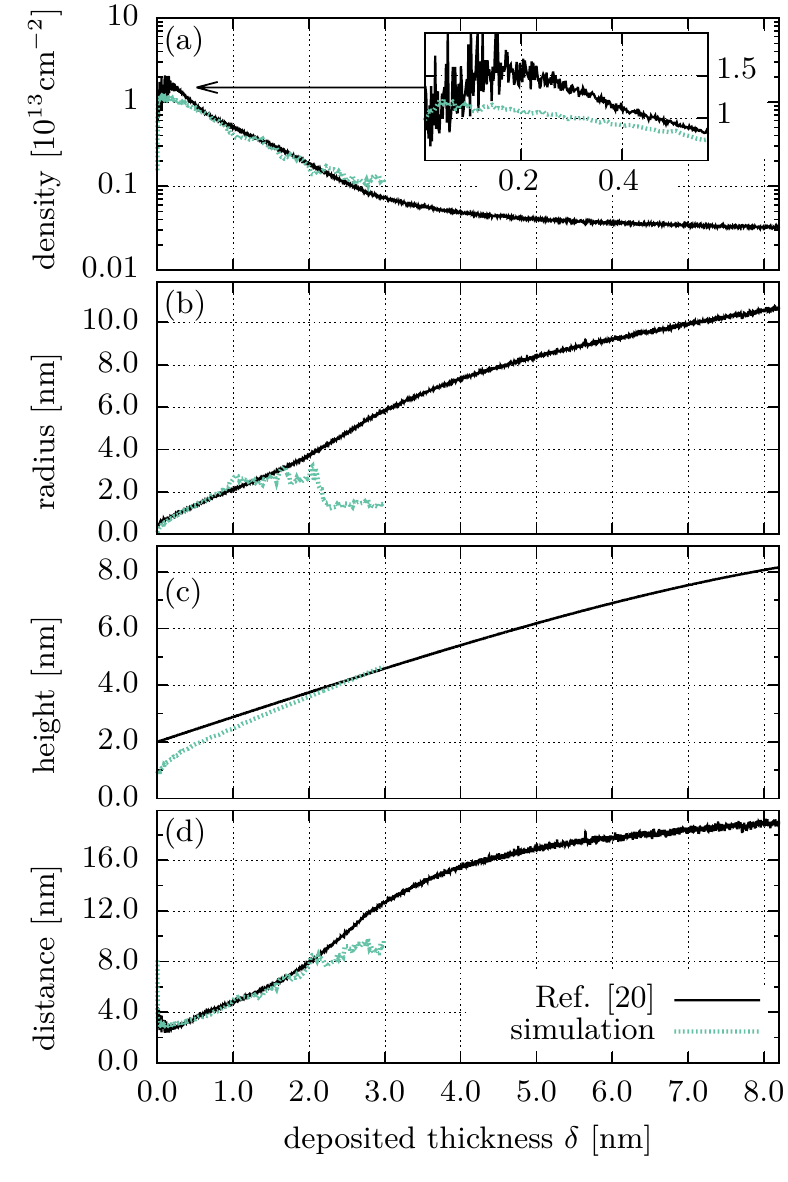}
\end{center}
\caption{Comparison of cluster properties
as obtained from experimental results in Ref.~\cite{Schwartzkopf2015}
and MD simulations with selected values for the
re-evaporation probability, $p_\mathrm{re}=\SI{1e-4}{}$,
and
the fraction of trapped Au particles, $\gamma = \SI{1e-2}{}$.
For deposited thicknesses $\delta$ less than about \SI{2}{nm},
all quantities agree with experimental trends.
However,
with the formation of a percolated network of non-spherical
clusters for larger $\delta  \gtrsim \SI{2}{nm} $, the cluster radius and the cluster distance
are no longer meaningful quantities.
}
\label{fig:long_run}
\end{figure}

\begin{figure*}
\begin{center}
\includegraphics[scale=1]{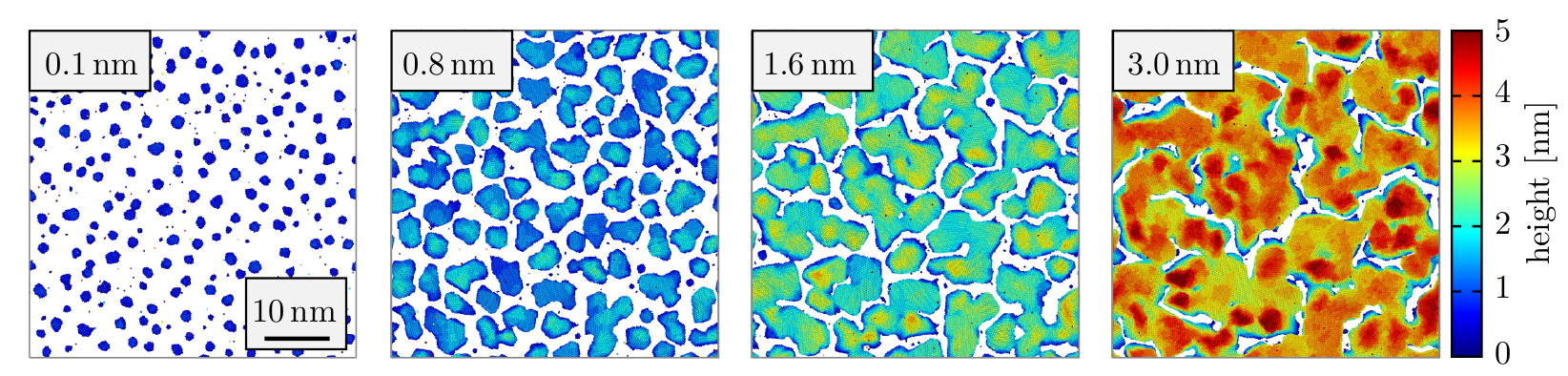}
\end{center}
\caption{Top view snapshots of the cluster morphology over a wide range of effective
deposited thicknesses from \SI{0.1}{nm} (\SI{18000}{} atoms) to \SI{3}{nm} (\SI{357000}{} atoms). The figure illustrates
the data set
 of a single run with the re-evaporation probability $p_\mathrm{re}=\SI{1e-4}{}$,
the fraction of trapped atoms $\gamma = \SI{1e-2}{}$ and the surface length \SI{45}{nm}.
The colors indicate the vertical position
of the atoms.
}
\label{fig:topview_snapshots}
\end{figure*}

\subsection{General behavior of the system}
\label{sec:generalComparison}
We start the presentation with a general
comparison of experimental results from Ref.~\cite{Schwartzkopf2015}
and selected simulation results of a run with a
computation time of several days on a computer cluster with $\sim$1000 CPU cores.
The corresponding simulation parameters are $p_\mathrm{re}=\SI{1e-4}{}$ for the re-evaporation probability
and $\gamma = \SI{1e-2}{}$ for the fraction of trapped Au particles.
In Fig.~\ref{fig:long_run}, the corresponding cluster densities, radii, heights and distances
are shown
over the full range of experimentally obtained effective film thicknesses $\delta$.
Whereas in the experiments, a film with the thickness $\delta=\SI{8}{nm}$
was deposited,  we restricted the longest simulation
run and most of the other runs to the deposition of a film with the thickness $\delta=\SI{3}{nm}$
and $\delta\approx\SI{1}{nm}$, respectively.
These restrictions of simulated thicknesses are imposed by
the computational run time which scales according to $\mathcal{O}(N_\mathrm{atoms})$.
With the given surface size, each deposited nanometer
corresponds to \SI{1.2E5}{} atoms. The simulation time depends on the values of $\gamma$
and $p_\mathrm{re}$, because these parameters affect the sticking of atoms
to the bare surface.
While only \SI{7.2E7}{} steps (\SI{72}{ns}) are required to deposit the first nanometer
with the parameters $\gamma = \SI{5E-2}{}$ and $p_\mathrm{re}=\SI{1E-4}{}$,
it takes \SI{18.7E7}{} steps (\SI{187}{ns}) to deposit the same
thickness with the parameters $\gamma = \SI{1E-3}{}$ and $p_\mathrm{re}=\SI{1E-1}{}$.

Although only the initial period of the film growth in strong non-equilibrium is simulated,
the data is sufficient capture the
non-monotonic behavior of the cluster density and the average cluster distance.
In the following, we describe in detail how the cluster properties evolve.
For a better understanding, we also refer to
the snapshots of simulated cluster structures in Fig.~\ref{fig:topview_snapshots},
which serve as an illustration to the results shown in Fig.~\ref{fig:long_run}.

The number density of clusters shown in Fig.~\ref{fig:long_run}(a)
exhibits a sharp maximum at $\delta\approx\SI{0.1}{nm}$ followed by
a nearly exponential decay that transitions into a saturated state
beginning at $\delta\approx\SI{5}{nm}$. The maximum originates from
two processes with opposing effects: while the addition of atoms to the system
leads to the formation of new clusters, other clusters coalesce and thus reduce the number of clusters.
In the case of small clusters, the coalescence is mainly driven by the diffusive motion
that the clusters perform along the surface.
However, the more atoms a cluster contains, the slower this motion becomes. In this case,
lateral aggregation is caused by the attachment of atoms and small clusters to existing clusters.
Once a large fraction of the polymer surface is covered with metal,
most incoming atoms are directly deposited onto an existing big cluster. Consequently,
lateral growth becomes less important with increasing film thickness.
This explains why the density only slowly changes for high values of $\delta$.
Comparing experimental data and simulation results,
we find partial quantitative agreement,
in particular concerning the $\delta$-value of the maximum
and the subsequent decaying values of the density.
As we will show in the discussion below,
qualitative agreement is found for many parameter sets, but -- 
especially near the position of the maximum density --
the numerical data is very sensitive to simulation parameters.

In Fig.~\ref{fig:long_run}(b), the cluster radii can be compared.
While the experimental curve does not display any deviation from the
monotonic growth behavior, the simulation results start fluctuating
for $\delta > \SI{1}{nm}$. This behavior is attributed to the aforementioned
non-spherical cluster shapes and poor statistics due to the finite size of the simulation box.
For $\delta > \SI{2}{nm}$, there is even a drop of the curve.
The reason for this is the co-existence of few large cluster structures
and several very small clusters that all contribute with the same weight
in the averaging procedure to calculate the mean radius.
These fluctuations do not occur in the experimental data, because
the monodisperse cluster model is also used for elongated cluster structures.
Consequently, the values we obtained for the radii are only meaningful
for thicknesses up to $\delta \approx \SI{1}{nm}$. In this regime, 
we find indeed good agreement with experimental data.

Similar to the cluster radii, the cluster heights shown in Fig.~\ref{fig:long_run}(c)
are also characterized by a monotonic growth. 
Since the film becomes more dense with increasing effective thickness $\delta$,
the cluster heights approach the same values as the film thickness for large $\delta>\SI{2.5}{nm}$.
For small $\delta$, 
all simulated heights are significantly smaller than the
experimental values. For all investigated parameter sets,
the difference is about a
factor of 2 for very small $\delta$.
One reason for this discrepancy might the fact that in the experiment
the heights of very thin films
were not directly measured, but obtained from an extrapolation
of the values for larger $\delta$.
\begin{figure*}
\begin{center}
\includegraphics[scale=0.98]{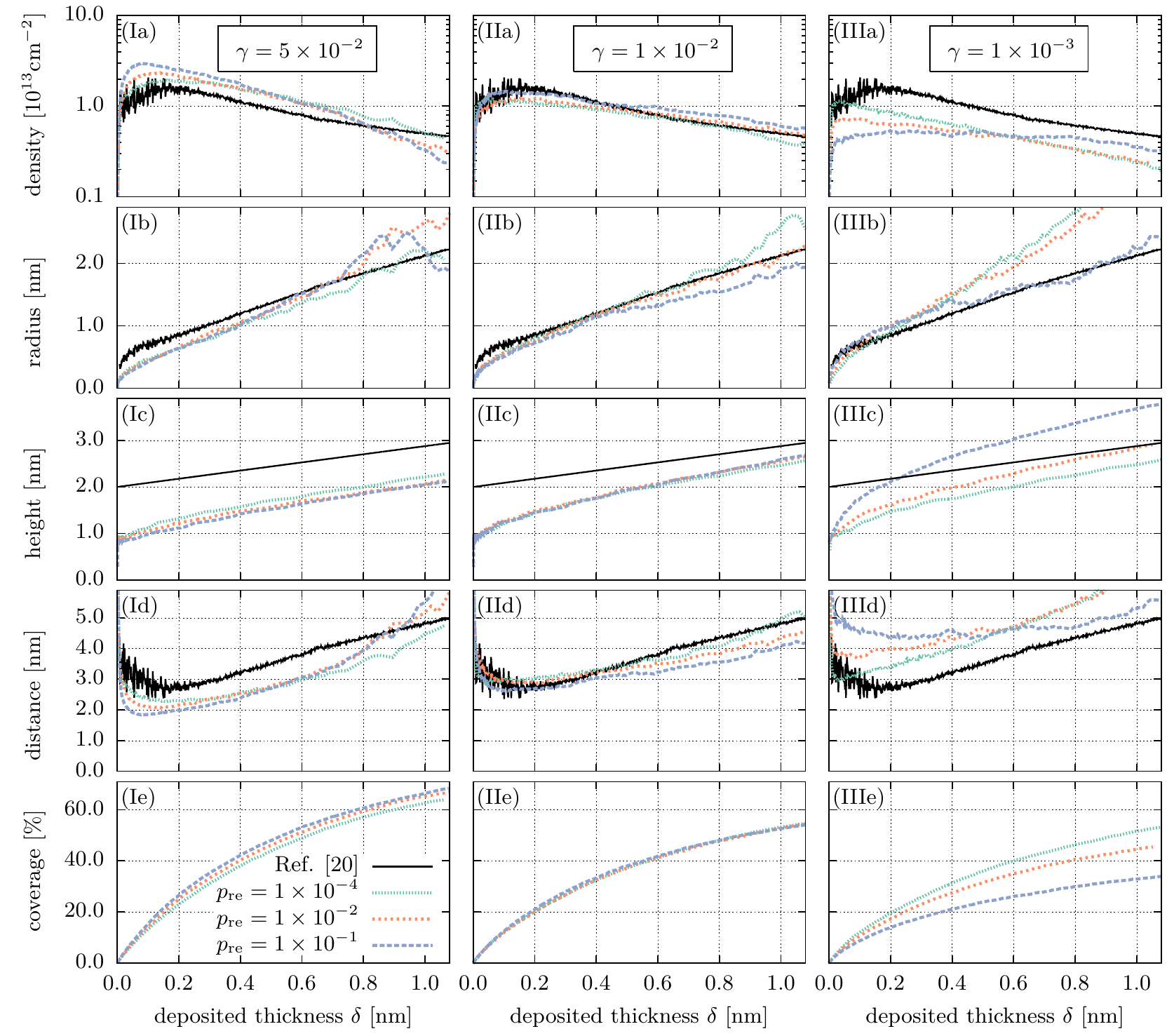}
\end{center}
\caption{Evolution of cluster properties (rows 1-4)
and surface coverage (row 5) during the deposition process
for different fractions
of trapped Au atoms $\gamma$ (represented by each column)
and different re-evaporation probabilities $p_\mathrm{re}$
(represented by line styles). The black solid curves
show the experimental data from Ref.~\cite{Schwartzkopf2015}.
}
\label{fig:td_results}
\end{figure*}

Nevertheless, the origin of this discrepancy remains an open question
in this analysis. 
However, we find that the agreement is improved for larger film thicknesses.
Furthermore, we remark that the simulation data does not
exhibit any fluctuations in the regime of larger $\delta$ because
the histogram which is employed to determine the height becomes more accurate
the more particles are added to the system.

Finally, Fig.~\ref{fig:long_run}(d) allows us to
compare the distances of the clusters.
As can be expected, we find a minimum of the distances for small thicknesses $\delta$
that are characterized by a high number density of very small clusters.
Although the space between the clusters is decreased by increasing the film thickness
$\delta$, the values for the distances increase as they refer to the centers of masses
of the clusters.
Since in both the experimental and the computational data,
the distances and the densities are not determined independently,
the evolution of the mean cluster distance exhibits the inverse trends
of the density. In particular, the local extrema can be found
at the same $\delta$-values.
Before the statistical fluctuations set in
close to $\delta \approx \SI{2}{nm}$,
most simulated distances deviate by less than $\SI{5}{\%}$
from the experimental values. While the limit $\delta \to 0$
is not resolved by the experimental data,
the computational data
diverges due to the vanishing
cluster density.

\begin{figure*}
\begin{center}
\includegraphics[scale=0.999]{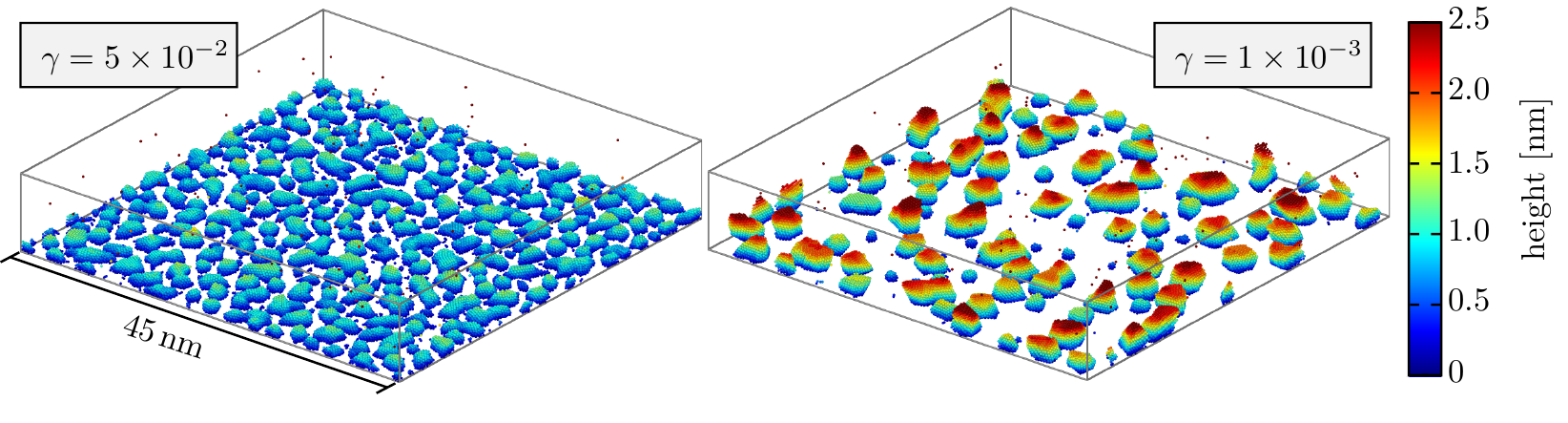}
\end{center}
\caption{Simulation snapshots
of the film morphologies
for different fractions of
trapped Au atoms $\gamma$, but equal
effective film thicknesses $\delta=\SI{0.46}{nm}$.
The re-evaporation probability is $p_\mathrm{re}=\SI{1e-1}{}$ for both cases.
}
\label{fig:comparison_defects_snapshots}
\end{figure*}

\subsection{Analysis of the influence of the model parameters}
\label{sec:parameter}
Despite the microscopic treatment of the clusters,
the implementation of re-evaporation and trapping processes
is only a rough approximation to the real behavior.
As a consequence, 
we conclude by discussing in detail how 
the corresponding parameters $p_\mathrm{re}$ and $\gamma$
affect the evolution of the deposited film.
For this analysis, we restrict the maximum thickness to
$\delta \approx \SI{1}{nm}$ to make sure that the values are meaningful.

In Fig.~\ref{fig:td_results}, the results of a parameter scan
are presented for three different values of $\gamma$ (columns I-III)
and three different values of $p_\mathrm{re}$ indicated by the line styles.
For further comparison, we also show the experimental data
and add results for the surface coverage which has not been investigated in
Ref.~\cite{Schwartzkopf2015}.
Since both the parameters $\gamma$ and $p_\mathrm{re}$
have an effect on the sticking of atoms to the surface, it is clear that
the influence of these parameters can only be understood
by considering both of them together.
For example, the influence of the trapping parameter $\gamma$
is relatively weak if the re-evaporation probability is small.
Hence, the $\gamma$-dependence of all curves with the parameter 
$p_\mathrm{re}=\SI{1e-4}{}$ is not as strong as for the corresponding curves
with $p_\mathrm{re}=\SI{1e-1}{}$.

A general influence of raising the trapping parameter $\gamma$
is a reduction of the number of diffusing particles, an increased number
of nucleation sites and a higher chance that
deposited atoms remain on the surface. Consequently,
the cluster density (row (a)) can be increased by increasing $\gamma$.
This can clearly be observed for the low re-evaporation
probability $p_\mathrm{re}=\SI{1e-1}{}$ near the maximum
values of the density. For further illustration of this trend, we also
provide simulation snapshots of films with fixed parameters
$\delta = \SI{0.46}{nm}$ and  $p_\mathrm{re}=\SI{1e-1}{}$, but different
$\gamma$-values in Fig.~\ref{fig:comparison_defects_snapshots}.
Depending on the value of $p_\mathrm{re}=\SI{1e-1}{}$,
$\gamma$ 
affects not only the value of the number density,
but also
the effective thickness $\delta$
at which the maximum occurs. For $p_\mathrm{re}=\SI{1e-4}{}$,
it is apparent that a reduction of trapped particles leads to faster
coalescence of clusters, which shifts the maximum to smaller values of $\delta$.
As a final observation on cluster densities,
we point out that there is a trend reversal of the influence of $p_\mathrm{re}$
in the third column. If the number of trapped particles
is low, the number of clusters on the surface is altered by
reducing the re-evaporation probability.
However, for $\gamma=\SI{5e-2}{}$ and $ \gamma=\SI{1e-2}{}$
the contrary behavior is found. This can be explained
by the fact the a high value of the re-evaporation probability
hampers the lateral growth and thus the coalescence of the numerous small clusters formed
at the defect sites.
The numerical data of the cluster radii and heights in rows (b) and (c) confirm
this trend. Apart from that, 
the dependence of these quantities
on $p_\mathrm{re}$ is weak in the early growth stage.
The only exception are the cluster heights for $\gamma=\SI{1e-3}{}$,
where a large value of the re-evaporation probability is associated with a
relatively big cluster height.
This is also illustrated in the right snapshot of Fig.~\ref{fig:comparison_defects_snapshots},
where the clusters are particularly high.
Since the cluster distance is computed from the number of clusters
according to Eq.~\eqref{eq:meanClusterDistance}, the values
for the distance in row~(d) of Fig.~\ref{fig:td_results} mirror all trends that
have been described for the density.
Finally, in row (e), we also provide data for the surface coverage, which
always increases monotonically with $\delta$ and mostly
exceeds $\SI{50}{\%}$ for $\delta = \SI{1}{nm}$.
Only in the case of a small trapping
parameter, $\gamma=\SI{1e-3}{}$, where the cluster densities are relatively
low, the re-evaporation probability has a noticeable effect on the surface coverage.
Although we cannot make a direct comparison with experimental data in this work,
we remark that similar experimental results have been reported
in Ref.~\cite{Schwartzkopf2013}, where, for example, a value of \SI{60}{\%}
has been measured for the surface coverage at the thickness $\delta = \SI{1}{nm}$. This value is close to the value
we obtained for $\gamma=\SI{1e-2}{}$ and $\gamma=\SI{5e-2}{}$.

Comparing the cluster properties again to the data from Ref.~\cite{Schwartzkopf2015},
we find qualitative agreement for most of the data curves.
The best quantitative agreement is found for $\gamma=\SI{1e-2}{}$, although none
of the parameter combinations removes all discrepancies of the cluster heights or maximum cluster densities
at once.

\begin{figure}
\begin{center}
\includegraphics[scale=1]{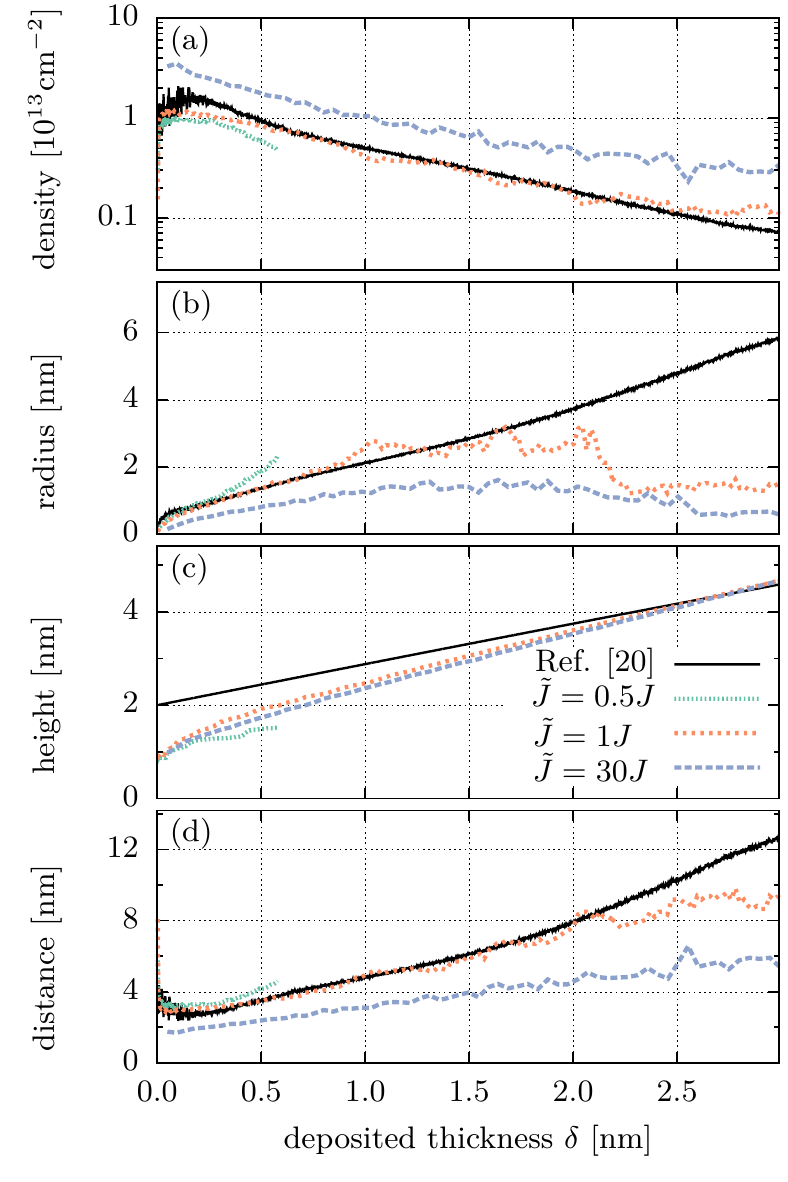}
\end{center}
\caption{Cluster properties for different deposition rates
$\tilde{J} = s \cdot J$, where the case $s=1$ corresponds to the deposition
rate that has been used for all other results in this work.
The results have been obtained with the
re-evaporation probability $p_\mathrm{re}=\SI{1e-4}{}$
and
the fraction of trapped Au particles $\gamma = \SI{1e-2}{}$.
For comparison, the experimental data from Ref.~\cite{Schwartzkopf2015}
(also corresponding to $s=1$)
are shown by the black solid lines.
}
\label{fig:depositionRate}
\end{figure}

\subsection{Influence of the deposition rate}
We conclude the investigation of simulation parameters by analyzing the
influence of changing the deposition rate.
In Fig.~\ref{fig:depositionRate}, results for the cluster properties
are shown for different deposition rates, which are expressed
as multiples of the rate $J$ that we used for the other simulations.
The simulations were performed with the parameters
$\gamma=\SI{1e-2}{}$ and $p_\mathrm{re}=\SI{1e-4}{}$,
which yielded the best agreement with experimental data in the above analysis.

The plot demonstrates that a high deposition rate, $\tilde{J}= 30J$
leads to higher cluster densities, but similar to what has been found in the investigation
of the previous section, the clusters have smaller radii, smaller heights and they are closer to each other.
The reason for the observed behavior is that a reduction of the cluster
number by coalescence is suppressed if the diffusive motion
is slow compared to the deposition process.
So far, our simulation results do not allow to predict whether these
differences will vanish for larger values of $\delta$.
Although only high rates are typically of interest for fast
processing in technological applications, we also show another
curve that represents the simulation with the low rate $\tilde{J}= 0.5J$.
Since the corresponding CPU time is roughly twice the time for $\tilde{J}= J$,
we only cover a small regime of effective thicknesses.
However, this is sufficient to observe trends that
are different from the behavior in the 
case of high rates. As the clusters have more time to move on the surface
between deposition events, they have more time to agglomerate. This results
in slightly reduced densities and bigger cluster radii. The cluster heights
are even smaller than for $\tilde{J}= 30J$, which indicates
that the heights scale non-linearly with $\tilde{J}$.
The occurrence of reduced cluster heights for small deposition rates
can be attributed to relatively long equilibration times in which
the cluster radii grow while their heights shrink.

\section{Conclusions and Outlook}
In this work, we presented a simulation scheme
that can be used to simulate the cluster growth process
of sputter-deposited metal atoms on a disordered, 
fluid-like surface, e.\,g., a polymer substrate.
The simulation model takes advantage of some ideas from previous kinetic Monte Carlo
models that have 
been successfully applied to simulate similar systems \cite{rosenthal13, Abraham15}.
These ideas comprise the representation of the substrate
as a continuum, which causes random walks of the metal particles,
as well as the implementation of simplistic processes
that take into account the desorption of atoms on the surface
and the creation of defects at the surface by the impingement of highly energetic
particles from the plasma.
However, while the KMC models rely on simple growth
models that neglect atomistic details of the clusters,
the new aspect of the present model is the treatment of each individual metal atom
with molecular dynamics.
Such a microscopic approach allows one to simulate
the formation of metallic structures
with complicated geometries, which is impossible with 
the KMC models that
require that only spherical or columnar shapes occur.

Compared to KMC simulations, the drawback of this new approach
is the limitation of accessible system sizes.
While length scales larger than \SI{1}{$\mu$m}
and time scales in the range of minutes and hours can be accessed with KMC,
the scope of this model is restricted to tens of nanometers
and the influence of large time scales can only be imitated
by a strong proportional
increase of the deposition rate and the diffusion coefficient.
In the simulations, these quantities are calculated according to Eqs.~\eqref{eq:diffusioncoeff} and \eqref{eq:ratio_xi}
after fixing the damping parameter.
Although the resulting values are up to 9
orders of magnitude larger than typical experimental values,
the observed nucleation and coalescence
of stable clusters is in accordance with the expected physical behavior
and the wide-ranging agreement with experimental data for gold
on polystyrene from Ref.~\cite{Schwartzkopf2015} 
supports the significance of the results.
Nevertheless, it presently remains difficult to perform benchmarks
that go beyond comparisons with experimental results.
However, we can state that an upper limit for the damping parameter is given by the value
for which the resulting time between the deposition
of two atoms is no longer big enough to prevent the atoms from forming
clusters above the boundary $z_\mathrm{max}$. For the parameter set we used,
only atomic deposition with subsequent cluster formation on the surface
is observed.

We used the model to investigate how the trapping
of metal atoms, the re-evaporation of atoms from the surface and the deposition rate
affect the evolution of the cluster morphology. Especially in the early
stage of the deposition, specific choices of these parameters may lead to different
amounts of mobile surface atoms, which are necessary to trigger cluster coalescence.
Consequently, depending on the parameters, one often observes either high densities
of clusters with small radii or lower densities of clusters with larger sizes.
It can be expected that the influence of the trapping and the re-evaporation becomes
less dominant in the case of thick films with high surface coverage, but currently,
the simulated film thicknesses are not sufficient to quantify that statement.
Reliable statements about the horizontal cluster dimensions
can be made as long as most of the clusters 
do not extend over the periodic boundaries. For the present results,
which were obtained with a simulated surface area of
$\SI{45}{nm} \times \SI{45}{nm}$, this is the case for effective film thicknesses
up to $\delta \approx \SI{1.5}{nm}$

While the system sizes we investigated in this work are already sufficient to investigate
the nucleation and coalescence of clusters,
the significance of this type of simulation can 
be easily enhanced just by spending more computation time on the simulation of 
larger segments of the surface and thicker films.
Furthermore,
a rather comprehensive improvement of the simulation model
could include the resolution of individual plasma species, e.\,g., metal and gas ions,
charging of clusters, e.\,g., by using modified potentials.
In particular, experimental or computational studies that yield the fluxes
and associated energy distributions of all particle fluxes towards the surface
would be of interest to improve the simulations.
Another potential application of the simulation scheme
is the simulation of the co-deposition processes of, for example, metal and polymer,
which could be realized by
implementing a continuous shift of the surface.
Similar implementations have already been applied in the framework of the 
aforementioned kinetic Monte Carlo
simulations \cite{rosenthal13, Abraham15}.

\begin{figure}
\begin{center}
\includegraphics[scale=1.0]{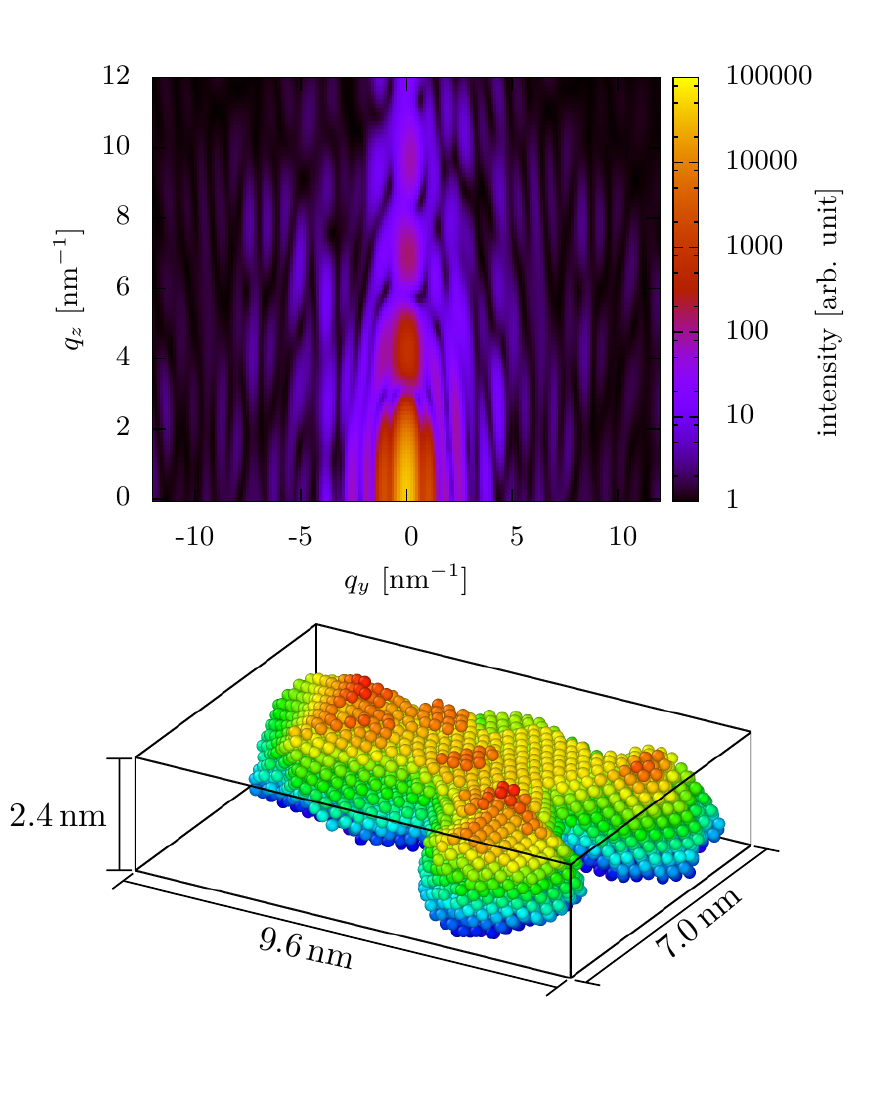}
\end{center}
\vspace{-1cm}
\caption{Scattered intensity (top)
of an exemplary cluster (bottom) that is contained
in a film with the effective thickness $\delta = \SI{1.1}{nm}$.
}
\label{fig:formFactor}
\end{figure}

Finally, we mention that it would also be of interest to simulate
the scattering of X-rays at the simulated cluster structures, because
this would permit a one-to-one comparison with the quantities that are actually
measured in the experiments. Although this is no fundamental problem \cite{Lazzari2002},
so far, the surface sizes of the performed simulations are not big enough
to generate data with sufficiently small statistical errors.
In particular, the interference effects that dominate
during the early stages of cluster growth (see Sec.~\ref{sec:exp_method})
cannot be resolved.
Yet, even with the existing data, 
there is another potential application in the field of computational
evaluations of GISAXS experiments.
As has been described in Sec.~\ref{sec:exp_method}, the real-space data of cluster
structures is often obtained by adjusting the properties of simple geometrical
objects so that the scattered intensity of the model system fits the measured intensity
obtained in the
GISAXS measurement. 
Possibly, one could achieve better fit results
by performing the evaluations with more realistic cluster
shapes instead of, e.\,g., hemispheres, cylinders or cubes. For instance, the
software \textit{BornAgain} \cite{bornagain}, which is currently being developed,
facilitates this, because it
includes the functionality to perform the evaluation of GISAXS data
with custom cluster shapes. The required input, which could be provided
by simulation results, is the Fourier transform
\begin{equation}
F(\mathbf{q}) =  \int_S \exp(-\mathrm{i} \mathbf{q} \mathbf{r}) \, \mathrm{d}^3 r 
\end{equation}
of each cluster shape $S$ that shall be included. The inverse length
$\mathbf{q}$ corresponds to the scattering wave vector.
For practical
calculations of $F$, $S$ can be approximated by a set of cuboids.
For example we show the scattering features of a single cluster
with an irregular shape in
Fig.~\ref{fig:formFactor}, 
where the scattering cross section
has been calculated in simple Born approximation \cite{Lazzari2002}, according to
\begin{equation}
\frac{\mathrm{d}\sigma}{\mathrm{d}\Omega} (\mathbf{q}) = F(\mathbf{q}) F^*(\mathbf{q}) \;.
\end{equation}

\section{Acknowledgments}
This work is supported by the Deutsche Forschungsgemeinschaft
via SFB-TR 24 (project A5). We thank Alexander Hinz for
discussions of the results
and improvements to the manuscript.


%

\end{document}